\documentstyle{article}
\font\gotb eufm10 scaled \magstep1
\newcommand {\bb} {\bibitem}
\newcommand {\cc} {\cite}
\newcommand {\lll} {\lambda}
\newcommand {\lt} {\left}
\newcommand {\rt} {\right}
\newcommand {\vp} {\varphi}
\newcommand {\R} {\hat R}
\newcommand {\I} {\hat I}
\newcommand {\Q} {\hat Q}
\newcommand {\A} {\hat A}
\newcommand {\B} {\hat B}
\newcommand {\SS} {\hat S}
\newcommand {\PP} {\hat P}
\newcommand {\aA} {\tilde A}
\newcommand {\AAA} {\hbox {\gotb A}}
\newcommand {\QQ} {\hbox {\gotb Q}}
\newcommand {\BB} {\hbox {\gotb B}}
\newcommand {\sss} {\sigma}
\newcommand {\al} {\alpha}
\newcommand {\pP} {P (\aA)}
\newcommand {\bea} {\begin {eqnarray} \label}
\newcommand {\eeq} {\end {equation}}
\newcommand {\beq} {\begin {equation} \label}
\newcommand {\eea} {\end {eqnarray}}
\newcommand {\nn} {\\ \nonumber}
\newcommand{\rr}[1]{(\ref{#1})}

\begin{document}
\begin{center}
{\Large \bf Algebraic-statistical approach to quantum mechanics }\\

\vspace{4mm}

D.A.Slavnov\\
Department of Physics, Moscow State University,\\
 Moscow 119899, Russia. E-mail: slavnov@theory2.npi.msu.su \\

\end{center}

\begin{abstract}

It is proposed the scheme of quantum mechanics, in which a Hilbert space and the linear operators are not primary elements of the theory. Instead of it certain variant of the algebraic approach is considered. The elements of noncommutative algebra (observables) and the nonlinear functionals on this algebra (physical states) are used as the primary constituents. The functionals associate with results of a particular measurement.

It is suggested to  consider  certain ensembles of the physical states as quantum states of the standart quantum mechanics. It is shown that in such scheme the mathematical formalism of the standart quantum mechanics can be reproduced completely.
\end {abstract}

The majority of physicists has long hardened to the view  that in the famous dispute of Bohr with Einstein about bases of quantum mechanics Bohr was right. Nevertheless, in proposed paper the attempt is done to advance in the direction, which was defended by Einstein~\cc{ein}. The basic thesis of Einstein was that: in the modern form the quantum theory is the quite correct statistical theory of ensembles of physical systems but it is not a theory of elementary processes. In this sense it is not a complete theory. See also~\cc{epr} up over it.

The standart quantum mechanics (quantum mechanics of Bohr, Heisenberg, Dirac, von Neumann) is similar to thermodynamics in some sort. These sections of physics give the correct description of quantities averaged over the large ensembles of physical objects. In both cases it is possible to formulate postulates only for the ensembles (average quantities). On the basis of these postulates it is possible to construct the adequate theories.

However, it is known so far as it is a fruitful idea to consider the basic thermodynamic quantities as statistical the averages of more elementary quantities. Correspondingly it is very useful to consider that the laws of thermodynamics is not postulates but they are the consequences of the laws of statistical physics, which is grounded on probability theory and  complex of the laws driving behaviour of elementary physical objects.

In the proposed paper the attempt is done to realize something similar in case of quantum mechanics. However, in contrast to of classical thermodynamics, here it is not supposed that the laws for elementary physical objects are identical to the laws of classical mechanics.

Instead of this statement we shall accept much weaker postulate of the algebraic approach to quantum theory~\cc{emch}: the elements of some algebra correspond to observable quantities. This phrase denotes only that observable quantities can be  summed up, multiplied together, and multiplied by numbers. As all observable quantities are real, probably, it is possible to restrict oneself to a real algebra. The Jordan algebra of observables  (see~\cc{emch}) can set up a claim to this role. However, in this algebra the operation of multiplication is defined in a rather complicated manner, and there is no property of associativity. In this connection it seems to me justified to leave the framework of directly observable quantities and to consider their complex combinations, which further will be referred to as dynamical quantities.

Correspondingly we shall accept {\it the first postulate}. \\
 {\it To dynamical quantities there correspond elements of an involutive associative (generally) noncommutative algebra~\AAA, obeying: (i) for each element $ \R\in\AAA $ there is the hermitian element $\A$ $(\A^*=\A)$ that $\R^*\R = \A^2$; (ii) if $\R^*\R=0$, then $\R=0$}.

Here and everywhere further it is supposed that the requirement of presence of the unity element~$ \I $ is included in definition of  algebra. The hermitian elements of algebra~\AAA correspond to observable quantities. The set of these elements will be designated~$ \AAA_+ $. We shall consider that all elements of the real linear space~$ \AAA_+ $ correspond to observables. Strictly speaking, it is not always valid since rules of superselection can exist. But we shall not consider this problem in the present paper.

We shall borrow {\it the second postulate} from the standart quantum mechanics. \\
{ \it Let be $ [\A, \B] \ne0 $. Then a measurement of one of the observables (at least one) causes uncontrollable perturbation to the value of other observable}. \\
 It is usually used the expression ---the observables are not simultaneously measurable --- though this expression does not describe the phenomenon quite exactly.

In connection with this postulate commutative subalgebras of the algebra~\AAA will play essential role in the further. Let us designate a maximal real commutative subalgebra of the algebra~\AAA by~\QQ {} ($ \QQ\equiv \{\Q \} \in\AAA_+ $). It is a algebra of the simultaneously measurable observables. If the algebra~\AAA {} is commutative (algebra of classical dynamical quantities), there is one such subalgebra. If the algebra~\AAA {} is noncommutative (algebra of quantum dynamical quantities), there a lot of such subalgebras. Moreover, the set of the subalgebras~\QQ{} has potency of continuum. Really, already for algebra~\AAA {} with two not commuting hermitian generators $ \A_1 $ and $ \A_2 $ the real subalgebra $ \QQ_{\al} $ with generator $ \B (\al) = \A_1\cos\al + \A_2\sin\al $ will be algebra of type~\QQ {} for everyone~$ \al $.

  I remind of the definition of a spectrum $ \sss (\A; \AAA) $ of algebra~\AAA. The number $ \lll $ is a point of the spectrum of an element~$ \A $ iff the element $ \A-\lll\I $ has no an inverse one in the algebra~\AAA.

Generally speaking, the same element can have different spectrums in an algebra and in its subalgebra. However, for algebras of type~\QQ {} the statement is valid (see~\cc{rud}): $ \sss (\Q; \QQ) = \sss (\Q; \AAA) $ for everyone $ \Q\in\QQ $.

The hermitian elements of the algebra~\AAA {} are latent form of the observable quantities. The explicit form of the observable should be some number. In other words, for the representation of the explicit form of observables on hermitian elements of the algebra~\AAA {} some functional $ \vp $ should be set: $ \vp (\A) =A $ is a real number.The physically, the latent form of an observable $ \A $ becomes explicit as a result measurements. It means that the functional $ \vp (\A) $ gives the value of the observable $ \A $ which is obtained in a {\it concrete (individual)} measurement. We shall call this functional the physical state of the considered quantum object.

In an individual experiment the mutually commuting elements can be measured only. The sum and the product of observed data should correspond to the sum and the product of observables: $ \A_1 + \A_2 \to A_1+A_2 $, $ \A_1\A_2 \to A_1A_2 $.

Let us use the following definition (see \cc{rud}). Let \BB {} be complex (real) commutative algebra and $ \vp $ is a linear functional on~\BB. If
\beq {1}
\vp (\B_1\B_2) = \vp (\B_1) \vp (\B_2)
\eeq
for all $ \B_1\in\BB $ and $ \B_2\in\BB $, then the functional $ \vp $ refers to as by a complex (real) homomorphism on algebra~\BB. The functional, satisfying to equality~\rr{1}, also refers to as by a multiplicative  functional.

Let us accept {\it the third postulate}.

 {\it The physical state of a concrete quantum object is defined by the functional $ \vp (\A) \qquad (\A\in\AAA_+) $, for which restriction on any subalgebra of type~\QQ {} is a real homomorphism}.

 The functional $ \vp (\A) $ has properties (see~\cc{rud}):
\bea {2}
            &/1/& \vp (0) =0; \nn {}
            &/2/& \vp (\I) =1; \nn {}
            &/3/& \vp (\A^2) \ge 0; \nn {}
            &/4/& \mbox {if } \lll = \vp (\A), \mbox { then } \lll\in\sss(\A; \AAA); \nn {}
            &/5/& \mbox {if } \lll\in\sss (\A; \AAA), \mbox { then } \lll = \vp (\A) \mbox { for some } \vp (\A).
\eea
All these properties perfectly agree with results of individual experiments.

It seems natural additionally to require existence of an experiment (i.e. a functional~$ \vp (\;) $), which would separate various observables. In other words, it is valid {\it the fourth postulate}

$$
\vp (\A_1) = \vp (\A_2) \mbox { {\it for all} } \vp, \mbox { {\it iff} } \A_1 = \A_2.
$$

I note that in the standart quantum mechanics the more strong supposition is done: the observables coincide if all their average values coincide. In the standart quantum mechanics it is postulated also that the result of an individual measurement of an observable belongs to its spectrum. In the proposed approach it is obtained automatically (formula \rr{2}: /4/ and /5 /).

Only restriction of the functional $ \vp (\;) $ on the subalgebras~\QQ has property of  linearity , but  the physical states are not linear functionals on the set~$ \AAA_+ $ .

Set of all physical states of the considered physical system can be called by phase space of this system. That is, a point of the phase space $M $ is a (nonlinear) functional $ \vp $ on the set $ \AAA_+ $.

Just as in case of the classical phase space, the point in $M $ can be given with the help of a set of real numbers (coordinates). This set can be constructed as follows. We select some subalgebra $ \QQ_1 $ (of type \QQ). Let $G (\QQ_1) $ be a set of generators of the algebra $ \QQ_1 $. The functional $ \vp $ maps the set $G (\QQ_1) $ into the set $S_1 $ of real numbers (points of the spectrum of the algebra \AAA):
\beq {4}
G (\QQ_1) \stackrel {\vp} {\longrightarrow} S_1.
\eeq
Further we select another subalgebra $ \QQ_2 $. If $G (\QQ_1) \cap \QQ_2\equiv \tilde G_1\ne \emptyset $, the elements of the set $ \tilde G_1 $ we take as  subset of generators of the algebra $ \QQ_2 $. We expand this subset arbitrarily to have  a complete set $G (\QQ_2) $ of generators of the algebras $ \QQ_2 $. Similarly to the formula \rr{4}
$$
G (\QQ_2) \stackrel {\vp} {\longrightarrow} S_2.
$$
Then, we build generators of the subalgebras $ \QQ_3 $, including in their composition the elements of the set $ [G (\QQ_1) \cup G (\QQ_2)] \cap \QQ_3 $. And so on.

If the algebra \AAA {} is commutative (the classical physical system), the process will break at the first stage. And we shall obtain coordinates of the usual classical phase space. If the algebra \AAA {} is noncommutative, a potency of the set of the subalgebras $ \QQ_i $ is equal to the continuum. The process never will break. The set of the coordinates $ \cup_i S_i $ will be infinite. Moreover, the set of the coordinates will have a potency of the continuum. Each coordinate is a real point of the spectrum of the algebra~\AAA.

It is possible to see that the structure of a phase space of a quantum system is similar to the structure of a phase space of a classical real field. Therefore, Einstein ideas that one constituent of the base of the complete quantum theory should be classical fields of the general relativity, perhaps, are not absolutely motiveless.

Now we shall introduce a construction, which corresponds to pure state in the standart quantum mechanics. The functional $ \vp $ maps a set $ \QQ = \{\Q \} $ (maximal commutative subalgebra) into the set of real numbers:
$$
\{\Q \} \stackrel {\vp} {\longrightarrow} \{Q = \vp (\Q) \}.
$$
For different functionals $ \vp_i (\;) $, $ \vp_j (\;) $ the sets $\{\vp_i (\Q) \} $, $ \{\vp_j (\Q) \}$ can either differ or coincide. If for all $ \Q \in \{\Q \} $ is valid $ \vp_i (\Q) = \vp_j (\Q) =Q $, then we shall call physical functionals $ \vp_i (\;) $ and $ \vp_j (\;) $ as $ \{Q \} $-equivalent functionals. Let $ \{\vp \}_Q $ be the set of all $ \{Q \} $-equivalent functionals. Let us call the set $ \{\vp \}_Q $ a quantum state, and we shall designate it~$ \Psi_Q $.
 It is clear, that for unique fixing~$ \Psi_Q $ it is not necessary to fix the set of all observables $ \{\Q \} $. It is possible to restrict oneself to values of the functionals for generators of the algebras~\QQ.

To the quantum state $ \Psi_Q $ there corresponds a surface (infinite-dimensional, more precisely, the dimensionality has a potency of continuum) in phase space of the quantum system.

The physical state cannot be uniquely fixed. Really, in one experiment we can measure only mutually commuting observables. All these observables are elements of a certain set $ \{Q \} $. But all $ \{Q \} $-equivalent functionals have identical values on any observable $ \Q\in \{Q \} $. Therefore, with the help of the performed experiment we cannot distinguish one physical state from another if both of them belong to the set $ \{\vp \}_Q $.

The second experiment will not help, as, generally speaking, in this experiment we shall deal with other physical state. Anyway, we do not have any physical possibility to reproduce safely the physical state, which we dealt in the first experiment with.

The most strict fixing of the physical state $ \vp $, which can be realized practically, consists in assigning it to some set $ \{\vp \}_Q $, i.e. to a certain quantum state. For this purpose it is sufficient to perform measurements of mutually commuting observables, it is possible to restrict oneself to corresponding generators only. In principle it can be done in one experiment.

Strictly speaking, it is valid only, when there are no identical particles in the physical system. The fact is that a measuring device does not distinguish, what of identical particles has hit  it. Therefore, to keep validity of the statement of the previous paragraph, it is necessary to extend the definition of quantum state.

Let the physical system contain identical particles. Let $ \{\vp \}_Q $ be a set of $ \{Q \} $-equivalent functionals. We shall call a functional $ \tilde \vp $ weakly $ \{Q \} $-equivalent to the functionals $ \{\vp \}_Q $, if there is such set $ \{\Q ' \} $, on which the functional $ \tilde\vp $ has the same values, what functionals $ \{\vp \}_Q $ has on the set $ \{\Q \} $. The set $ \{\Q ' \} $ should be a map of the set $ \{\Q \} $, in which the observables, concerning to one of the identical particles, are substituted by corresponding observables, concerning to another.

Now we call the set of all weakly $ \{Q \} $-equivalent functionals a quantum state $ \Psi_Q $. Further let us use  label $ \{\vp \}_Q $ for the set of all weakly $ \{Q \} $-equivalent functionals, and we shall omit the word "weakly" at definition of the equivalence.

 Thus, in essence we can not have the complete information about a physical state of a concrete physical object. The maximal observable and controllable information on the physical object is concentrated in the quantum state.

Actually, it is possible to obtain the information about a membership of the physical state of the concrete object to the intersection of two quantum states (see~\cc{slav}). However, this information always concerns  the state, in which the object was in past and    left  not by completely controlled manner.

It follows from this words that we always have the inexact information about the initial physical state of a concrete quantum object. Therefore, even under the strictly deterministic laws of evolution we can do only probabilistic predictions for values of the majority of observables. The situation is quite similar to what takes place in statistical physics.

 The quantum phase space has potency of continuum. Beforehand we have no special possibilities to organize experiment so that  the preassigned physical state appeared in it. Therefore, probability of appearance of the investigating quantum object in a preassigned physical state is equal to zero. It follows from here, the probability is equal to zero that in two different experiments we shall meet two identical physical states. Moreover, in anyone finite or denumerable set of experiments the probability of meeting two identical physical states is equal  to zero. Certainly, it does not signify that such event is impossible. Nevertheless, we can consider that in different experiments we always deal with  different physical states.

We use  certain positions of probability theory. Let us consider a phase space (set of physical states) as a general set and each measurement of  observable as a trial (term of probability theory). We shall consider experiment, in which the measured value of the observable $ \A $ is no more $ \aA $, i.e. $ \vp (\A) =A\le\aA $, as event $ \aA $.

This event is not unconditioned. In quantum theory there is a specific constraint: one trial is not to be  event for measurement of two noncommuting observables. The probability of event $ \aA $ is determined by the structure of phase space and  mentioned constraint. Let this probability be equal $ \pP $.

Let us consider sample of $n $  mutually independent random trials. By definition, the probability of appearance of the event $ \aA $ in each of these trials is equal to $ \pP $. Let the event $ \aA $ occur in this sample $k_n $ times. Then, by definition, $k_n/n $ is a probability of appearance of the event $ \aA $ in the sample. Over the Bernoulli theorem (see~\cc{korn}) the quantity $k_n/n $ converges over probability to $ \pP $ as $n\to \infty $.

Thus, the quantity, to which $k_n/n $ at $n\to\infty $ converges over probability, is a probability of appearance of the event $ \aA $ in any denumerable sample of independent random trials. This probability is identical at all such samples and is equal to $ \pP $. It determines the probability measure $ \mu (\vp) \qquad (\vp (\A) \le \aA) $ on any such sample.

Let us use the Khintchine's theorem (see~\cc{korn}),  formulated in the terms, which are used in the present paper.

Let $A_1 = \vp_1 (\A), \dots, A_n = \vp_n (\A) $ be a sequence of values of an observable $ \A $ in the mentioned sample of physical states. Everyone $A_i $ has the same distribution $ (\pP) $ and the finite expectation $ <A> $. Then the aleatory variable $ \bar A_n = (A_1 + \dots+A_n) /n $ converges over probability to $ <A> $ as $n\to\infty $.

Thus,
\beq {6}
<A> = \mbox {P-}\lim_{n\to\infty} \frac{1}{n} \Big(\vp_1(\A) + \dots + \vp_n(\A) \Big) = \int d\mu(\vp) \, \vp(\A).
\eeq
In the right-hand side it is necessary to integrate over denumerable random sample.

Denumerable random sample, which contains all measurement of quantity of an observable $ \A $, we shall call a set of relevant states for $ \A $ and we shall designate $ [\vp]^{\A} $. Similarly, we shall define a set of relevant states for an observable $ \B $. Let $ [\A, \B] \ne 0 $, then the observables $ \A $ and $ \B $ should be measured in different experiments. As the sets $ [\vp]^{\A} $ and $ [\vp]^{\B} $ are denumerable, the probability of their intersection is equal to zero. Therefore, for relevant states the constraint is fulfilled automatically with probability, equal to unity.

All reasonings remain in force, if the events satisfy  a certain additional constraint. In particular, (it is of special interest): under test all  physical states  belong to a certain quantum state $ \Psi_Q = \{\vp \}_Q $.

In this case the formula \rr{6} will be rewritten as follows
\beq {7}
        \Psi_Q(\A)=\mbox{P-}\lim_{n\to\infty}\frac{1}{n} \Big(\vp_1 (\A) + \dots + \vp_n (\A) \Big) = \int_{\{\vp \}^{\A}_Q} d\mu_Q(\vp) \, \vp (\A).
\eeq
Here $ \{\vp \}^{\A}_Q = \{\vp \}_Q\cap [\vp]^{\A} $ \qquad $ \vp_i\in \{\vp \}^{\A}_Q $.

The formula \rr{7} defines a functional (quantum average) on the set $ \AAA_+ $.  The totality of quantum experiments uniquely testifies in favour of what we should accept {\it the fifth axiom}.

 {\it A functional $ \Psi_Q (\;) $ is linear on the set $ \AAA_+ $.} \\
 It means
$$
\Psi_Q (\A + \B) = \Psi_Q (\A) + \Psi_Q (\B) \mbox { and in that case, when } [\A, \B] \ne0.
$$

In the standart quantum mechanics the similar statement is a consequence of three postulates:

{ \it (j) to each observable $ \A $ there corresponds a linear operator $ {\cal A} $ in some Hilbert space;

(jj) if the operators $ {\cal A} $ and $ {\cal B} $ correspond to observables $ \A $ and $ \B $, the operator $ {\cal A} + {\cal B} $ (simultaneous measurability $ \A $ and $ \B $ is not necessary) corresponds to the observable $ \A + \B $;

(jjj) the average value of an observable is equal to expectation of the corresponding operator.}

Any one of these statements does not follow uniquely from the experimental facts. Other matter, they do not contradict the totality of the experimental data and allow to construct the perfectly working  mathematical formalism of the standart quantum mechanics, .

Any element $ \R $ of the algebra \AAA {} can uniquely be represented in form $ \R = \A+i\B $, where $ \A, \B\in\AAA_+ $. Therefore, it is possible to extend the functional $ \Psi_Q (\;) $ up to the linear functional on the algebra \AAA: $ \Psi_Q (\R) = \Psi_Q (\A) +i\Psi_Q (\B) $.

Let $ \R^*\R = \A^2\in \{\Q \} $ ($ \A\in\AAA_+ $). If $ \vp\in\{\vp\}_Q $, then $ \vp (\R^*\R) =A^2 $. Therefore,
 $$
       \lt.\Psi_Q(\R^*\R)=A^2=\vp(\A^2)\rt|_{\vp\in\{\vp\}^{\A}_Q}.
$$
 From here
$$
 \| \R \|^2 \equiv\sup_Q \Psi_Q (\R^*\R) = \sup_{\vp\in [\vp]^{\A}} \vp (\A^2) > 0, \mbox { if } \R\ne0.
$$
Since $ \Psi_Q (\;) $ is a positive linear functional, the Cauchy-Bunkyakovsky-Schwarz inequality is valid
 $$
\Psi_Q (\R^*\SS) \Psi_Q (\SS^*\R) \le \Psi_Q (\R^*\R) \Psi_Q (\SS^*\SS).
 $$
Therefore, (see \cc{emch}), for $ \| \R \| $ the postulates for norm of the element $ \R $ are fulfilled: $ \| \R + \SS \| \le \| \R \| + \| \SS \| $, $ \| \lll\R \| = |\lll |\| \R \| $, $ \| \R^* \| = \|\R \| $.

 The algebra~\AAA {} become Banach space after extension over norm $ \| \; \| $. Since $ \vp ([\A^2]^2) = [\vp (\A^2)]^2 $, then $ \| \R^*\R \| = \| \R \|^2 $, i.e. the algebra~\AAA {} is $C^* $-algebra.
 Taking into account the formula \rr{2} /2/, we conclude that $ \Psi_Q (\;)$ is a positive linear functional on  $C^* $-algebra, satisfying  the normalizing  condition $ \Psi_Q (I) =1 $. Therefore, according to the Gelfand-Naumark-Segal construction (see \cc{emch}), the functional $ \Psi_Q (\;) $ canonically generates a Hilbert space and the representation of the algebra~\AAA{} by linear operators in this Hilbert space. In other words, in the proposed approach it is possible to reproduce the mathematical formalism of the standart quantum mechanics completely.

At will, the previous reasonings can be viewed as an underpinning of the postulates of the standart quantum mechanics, necessary for construction of its formalism. I shall note that the postulates, accepted in the present paper, are related to experiment more direct, than postulates of the standart quantum mechanics. Such only mathematical notions as vectors of Hilbert space (wave functions) and functionals in Hilbert space are not the primary elements of the theory in the proposed approach. They arise only at the second stage. The primary elements are the observables and physical states, which are related directly to the results of the experiment. It is possible to consider that they are connected with the material structure of the investigated quantum object and do not depend on the observer (see~\cc{slav}).

As opposed to that, the quantum states $ \Psi_Q (\ ) $ have some subjective element. The matter is that one physical state can belong to different quantum states $ \{\vp \}_Q $ and $ \{\vp \}_P $. That is, $ \vp\in \{\vp\}_Q\cap \{\vp \}_P $, where the states $ \{\vp \}_Q $ are classified by values of the set $ \{\Q \} $ of mutually commuting observables $ \Q $, and $ \{\vp \}_P $ are classified by values of observables $ \PP\in \{\PP \} $. The observables $ \Q $ and $ \PP $ do not commute among themselves. Then depending on what set ($ \{\Q \} $ or $ \{\PP \} $) we shall choose for the classification, the physical state $ \vp $ will be referred either to the quantum state $ \{\vp \}_Q $ or to the quantum state $ \{\vp \}_P $. This subjective component in quantum state is the origin of the famous Einstein-Podolsky-Rosen paradox~\cc{epr}. The variant of experiment, proposed Bohm~\cc{bom} for demonstrating the paradox, looks as follows.

Let a spin-zero particle decay into two particles $A $ and $B $ with spins 1/2 which scatter at large distance. Let us measure a projection of spin onto the axis $z $ for the particle $A $. Let the result will be $S_z (A) $. Then, using the conservation law, we can state that for the particle $B $ the projection of spin onto the axis $z $ is equal $S_z (B) = -S_z (A) $ with absolute probability. Therefore we should consider that the particle $B $ will be in the quantum state with fixed projection of  spin onto axis $z $ ( $ S_z(B)=-S_z (A) $) after measurements of  spin of  the particle $A $. However, for the particle $A $ we could measure the projection of  spin onto  axis $x $. Let the result  be $S_x (A) $. Then we could assert that after of such measurement the particle $B $ will be in the quantum state with fixed projection of  spin $S_x (B) = -S_x (A) $.

Thus, by his wish the observer "drives" the particle $B $ either into the quantum state $ \{\vp \}_{-S_z (A)} $, or into the quantum state $ \{\vp \}_{-S_x (A)} $, without acting on the particle $B $ physically in any way.

The authors of the paper \cc{epr}  estimated that such situation indicates incompleteness of quantum mechanics, since the following requirements are broken: (k) every element of the physical reality must have a copy in the complete physical theory; (kk) if, without in any way disturbing a system, we can predict with certainty (i.e. with probability equal to unity) the value of a physical quantity, then there is an element of reality corresponding to this quantity.

Bohr \cc{bohr} declaimed against this statement. Bohr supposed that the quantum particles $A $ and $B $ can not be considered as separate realities even when they have ceased to interact. In this sense the quantum theory essentially differs from classical one. In the classical theory, if the object negligibly weakly interacts with other objects, it can be studied as a separate reality. In the Bohr quantum theory it is not so.

The approach to quantum mechanics, proposed in the present paper, does not contradict the requirements, formulated in the paper \cc{epr}. The copy, appearing in the requirement (k),   is the physical state of each particle. The reality, which figures in the requirement (kk), will exist, if the physical state of each quantum object is uniquely determinated by its material structure. Such guess well agrees with the proposed approach (see \cc{slav}).

Thus, in the proposed approach the paradox does not arise. At any way of measurement of spin of the particle $A $ the physical state of the particle $B $ (the objective reality) will be same $\vp\in\{\vp\}_{-S_z(A)}\cap\{\vp\}_{-S_x(A)}$. The various quantum states of the particle $B $ arise due to a subjective choice by the observer of a device for measurement of the observable $ \A $.

It is seen from this example that the physical state~$ \vp $ can be considered as a specific hidden parameter. On the one hand, a particular physical state~$ \vp $ corresponds to a concrete event. In this sense it is a parameter. On the other hand, the physical state cannot be fixed uniquely with the help of experiment; it is possible only to establish its membership in this or that ensemble (quantum state). In this sense~$ \vp $ is a hidden parameter.

After the works of von Neumann hidden parameters have got bad reputation in quantum mechanics. In the famous monography~\cc{von} von Neumann asserts that the models with hidden parameters are incompatible with the basic postulates of quantum mechanics.

As such basic postulates of quantum mechanics von Neumann accepts, in particular, the postulates (j) and (jj). From these postulates he has made a deduction that any quantum-mechanical ensemble has dispersion. That is, in this ensemble not all average squares of observables are equal to squares of average corresponding quantities. In the monography~\cc{von} it is proved that such ensemble cannot be decomposed into zero-dispersion subensembles, as they simply do not exist. In return, it follows from here that  introducing of hidden parameters is impossible .

However, in the proposed  approach the postulates (j) and (jj) are absent. The corresponding statements are valid not for observables (elements of algebra), but only for their representations, which are generated by the functionals of a special form (quantum average). The von Neumann's reasoning is invalid for a physical state. A physical state is zero-dispersion "subensemble" comprising one element. Its existence is not forbidden since "average" values of observables over such ensemble are defined by the {\it nonlinear} functional $ \vp (\;) $. By virtue of nonlinearity of the functional $ \vp (\;) $ the physical state is not related to any representation of algebra of observables. Therefore, on physical states the observables are not representable by linear operators on a Hilbert space and the postulates ($j $, $jj $, $jjj $) can not be formulated for them.

In fact, in the monography~\cc{von} it is shown that the linearity of a state is in the conflict with  causality and the hypothesis about the hidden parameters. Herefrom von Neumann has made a deduction that causality is absent at the microscopic level, and the causality occurs due to averaging over large number of noncausal events at the macroscopic level.

The approach, formulated in the present work, allows to solve the same conflict in the opposite way. It is possible to suppose that there is causality at the level of the single microscopic phenomenon and the linearity is absent. The linearity of the (quantum) state occurs due to averagings over quantum ensemble. The transition from the single phenomenon to quantum ensemble replaces the initial determinism by probabilistic interpretation.

Besides von Neumann's reasoning, there is one more argument, not less famous, against schemes with hidden parameters, it is Bell inequality~\cc{bel}. The Bell has proved that if in a quantum system there are hidden parameters, the particular combination of correlation functions satisfies the certain inequality. The experiment does not confirm such deduction, and it agrees with results of the standart quantum mechanics.
For the proof of the Bell inequality it is essential that the hidden parameters, appearing in the different correlation functions, have identical ranges.

In the proposed approach the analogs of these ranges are the sets of states, 
relevant for corresponding observables. In the different correlation 
functions there are the noncommuting observables. Therefore, probability of 
intersection of the corresponding sets of relevant states is equal to zero.  
It follows from here that the probability of validity of the Bell inequality 
is equal to zero in our case. In more detail this problem is surveyed in 
paper~\cc{slav1}.

At axiomatic construction of theory a rather difficult problem frequently exists, whether there are nontrivial models satisfying  the formulated postulates. It is known that in the axiomatic approach to quantum theory of the field this problem is very painful. However, in the case, considered in present paper, this problem is solved simply.

The standart quantum mechanics represents huge amount of such models. It is possible to consider any of them, for example, the harmonic oscillator. In this system it is known the Hilbert space of quantum states and the set $ \tilde {\AAA}_+ $ of the selfajoint linear operators which are the observables. Over linearity this set is extend up to a complex algebra $ \tilde {\AAA} $. This algebra satisfies  all postulates, which the algebra of dynamical quantities \AAA{} should satisfy.

The structure of the sets $ \tilde {\AAA} $ and $ \tilde {\AAA}_+ $ is completely known. Therefore, in principle, we can find all maximal real commutative subalgebras. In these subalgebras we shall choose the systems of generators, as at construction of the coordinates in the phase space. Further we shall find spectrums  of   all the generators. Over the points of these spectrums, using property \rr{2} /5/, we shall construct functionals $ \tilde\vp $. Its restriction on each maximal subalgebra must realize a real homomorphism. These functionals will have properties of the physical states $ \vp $. Thus, we can construct the set of all physical states for the given physical system.

To construct  probability measure on this space, more precisely, on its subspaces, corresponding to each quantum state $ \Psi_Q $, it is possible as follows. We shall calculate  probability to detect the value $A $ of the observable $ \A $ in the state $ \Psi_Q $ over formulas of the standart quantum mechanics, using corresponding vectors of the Hilbert space. The  Gelfand-Naumark-Segal construction  ensures validity of this operation. After that we shall calculate probability to detect all values of the observable $ \A $, which satisfy  the condition $A\le\aA $. This probability will be the probability measure $ \mu_Q (\vp) \qquad \Big (\vp (\A) \le\aA, \quad \vp\in \{\vp \}^{\A}_Q\Big) $.

In present paper the kinematic and statistical problems of  quantum mechanics are surveyed only. It has allowed to do without any suppositions about  form of the interaction in  quantum mechanics.

It is possible to incorporate in the proposed  scheme of  quantum mechanics the dynamics, which describes not only unitary temporal evolution of quantum ensembles (interaction of the second sort, over terminologies of  von Neumann \cc{von}), but also interaction with  measuring device (interaction of the first sort). Within the framework of this dynamics the wave-corpuscle dualism and collapse of a quantum state can be explained.

The variant of such dynamics is described in paper \cc{slav}. However for construction of such dynamics it is required  more detailed suppositions about the structure of the quantum object. These suppositions have considerably less general character, than what are made in present paper. They can seem by much more disputable.

\end{document}